\documentclass[aps,prb,twocolumn,showpacs,preprintnumbers,
superscriptaddress,amsmath,amssymb,longbibliography,floatfix]{revtex4-2}

\newcommand{\ket}[1]{|#1\rangle}

\newcommand{\be}{\begin{equation}}
\newcommand{\ee}{\end{equation}}
\newcommand{\bea}{\begin{eqnarray}}
\newcommand{\eea}{\end{eqnarray}}

\newcommand{\Trms}{T_{\rm rms}}
\newcommand{\e}{\varepsilon}
\newcommand{\w}{\omega}
\newcommand{\s}{\sigma}
\newcommand{\G}{\Gamma}
\newcommand{\up}{\uparrow}
\newcommand{\down}{\downarrow}

\usepackage{tabularx}
\usepackage[dvipsnames]{xcolor}

  \newcommand{\Sec}[1]{Sec.\,\ref{Sec:#1}}
  \newcommand{\App}[1]{App.\,\ref{App:#1}}

  \newcommand{\Eq}[1]{Eq.\,\eqref{Eq:#1}}

  \newcommand{\Fig}[1]{Fig.\,\ref{#1}}

\def\Trms{\ensuremath{T_{\rm rms}}}
\def\TK{\ensuremath{T_{\rm K}}}

\def\VK{\ensuremath{V_{\rm K}}}
\def\Vlin{\ensuremath{V_0}\xspace}
\usepackage{ulem}
\usepackage{verbatim} 
\usepackage{amsmath}
\usepackage{graphicx}
\usepackage{dcolumn}
\usepackage{bm}
\usepackage{xspace}

\definecolor{oucrimsonred}{rgb}{0.6, 0.0, 0.0}
\usepackage[colorlinks, citecolor={blue!50!black}, urlcolor={blue!50!black}, linkcolor={red!50!black}]{hyperref}

\begin{document}

\title{Nonequilibrium steady-state thermoelectrics
of Kondo-correlated quantum dots}

\author{Anand Manaparambil}
\email{anaman@amu.edu.pl}
\affiliation{Institute of Spintronics and Quantum Information, Faculty of Physics, 
Adam Mickiewicz University, Uniwersytetu Pozna\'nskiego 2, 61-614 Pozna\'n, Poland}

\author{Andreas Weichselbaum}
\affiliation{Department of Condensed Matter Physics and Materials Science,
Brookhaven National Laboratory, Upton, New York 11973-5000, USA}
\author{Jan von Delft}
\affiliation{Arnold Sommerfeld Center for Theoretical Physics,
    Center for NanoScience, and Munich Center for Quantum Science and Technology,
    Ludwig-Maximilians-Universit\"at M\"unchen, 80333 Munich, Germany}

\author{Ireneusz Weymann}
 \affiliation{Institute of Spintronics and Quantum Information, Faculty of Physics, 
Adam Mickiewicz University, Uniwersytetu Pozna\'nskiego 2, 61-614 Pozna\'n, Poland}
\date{\today}


\begin{abstract}

The transport across a Kondo-correlated quantum dot coupled to two leads with independent temperatures
and chemical potentials is 
studied using a controlled non-perturbative, and in this sense exact numeric treatment
based on a hybrid numerical
renormalization group combined with time-dependent density matrix renormalization group (NRG-tDMRG).
We find a peak in the conductance at finite
voltage bias vs. the temperature gradient
$\Delta T = T_R - T_L$ across left and right lead.
We then focus predominantly on zero voltage bias
but finite $\Delta T$ far beyond linear response.
We reveal the dependence 
of the characteristic zero-bias conductance on the individual lead temperatures. 
We find that the finite-$\Delta T$ data
behaves quantitatively similar
to
linear response
with an effective equilibrium temperature
derived 
from the different
lead temperatures.
The regime of sign changes in the Seebeck coefficient,
signaling the presence of Kondo correlations, and its dependence on the individual lead temperatures provide 
a complete picture of the Kondo regime in the presence of finite
temperature gradients. 
The results from the zero-bias conductance and Seebeck coefficient studies
unveil an approximate `Kondo
circle' in the $T_L/T_R$
plane as the regime within
which the Kondo correlations dominate.
We also study the heat current
and the corresponding heat conductance
vs. finite $\Delta T$.
We provide a polynomial fit for our numerical results
for the thermocurrent as a function of the individual
lead temperatures which may
be used to fit experimental data in the Kondo regime.

\end{abstract}

\maketitle

\section{\label{Sec:Intro}Introduction}

Strong electronic correlations in a magnetic impurity coupled to electronic reservoirs
result in a many-body screening phenomenon, mediated by the conduction band electrons, known as the Kondo effect \cite{Hewson_1993}.
The Kondo effect manifests itself in the density of states of the impurity as a  narrow resonance peak around the Fermi level widely known as the Kondo-Abrikosov-Suhl resonance \cite{Suhl1965Apr,Abrikosov1970Nov}. 
This Kondo resonance that increases the low temperature
resistivity of bulk metal alloys \cite {deHaas1934May} has been found to be present
in various classes of nanostructures, involving single electron transistors 
\cite{Goldhaber-Gordon1998Jan,Cronenwett1998Jul,Schmid1998Dec, Goldhaber-Gordon1998Dec, Simmel1999Jul, Quay2007Dec}, 
nanowires \cite{ Jespersen2006Dec, Csonka2008Nov, Nilsson2009Sep, Kretinin2010Sep},
carbon nanotubes \cite{Nygard2000Nov,Kretinin2011Dec}, 
molecular magnets \cite{Yu2004Jan, Gonzalez2008Aug, Scott2009Apr}, 
adatoms \cite{Knorr2002Feb, Ternes2008Dec, Ren2014Jul}
and other quantum impurity systems \cite{Chen2011Jul, Riegger2018Apr,Seiro2018Aug, Shankar2023Jun,  vanEfferen2024Jan}.
Such nanostructures are very tunable and act as a robust platform
to explore various aspects of the Kondo effect \cite{Cronenwett1998Jul, Kouwenhoven2001Jan}. 
Moreover, the transport properties of Kondo-correlated impurity systems carry characteristic
signatures of the Kondo effect, which emerge at low temperatures near the Kondo energy scale.
Particularly, the zero-bias peak in the differential conductance \cite{Goldhaber-Gordon1998Jan,Cronenwett1998Jul}
and a sign change in the Seebeck coefficient at low energies \cite{Costi2010Jun, Svilans2018Nov,Dutta2019Jan, Gehring2021Apr,Hsu2022Apr} signify the presence of Kondo correlations in the system.
The characteristic density of states present in the quantum dots
makes them a class of prospective systems to work as efficient
energy-harvesters \cite{Hicks1993May, Mahan1996Jul, Heremans2004Sep, Szczech2011Mar, Heremans2013Jul,  Hershfield2013Aug, Sanchez2014Nov, Benenti2017Jun,Sanchez2016}.
Various proposals, such as the charge Kondo effect \cite{Andergassen2011Dec},
multi-quantum dot setups \cite{Gomez-Silva2018Feb,Santamaria2024Aug},
including the case of asymmetric couplings to the leads
\cite{PerezDaroca2018Apr,Manaparambil2023Feb}, have pointed towards
a considerable thermoelectric efficiency of quantum-dot based heat engines.
Thermoelectric quantum dot devices have also demonstrated promising
applications in sensing \cite{Dorda2016Dec,PerezDaroca2018Apr} and cooling technologies \cite{Giazotto2006Mar, Hwang2023Jun, Zeng2024Mar}.

An accurate description of the Kondo effect relies on the exact treatment of electronic correlations at low energy scales. 
Though many theoretical methods, including the Bethe-Ansatz \cite{Andrei1983Apr, Wiegmann1981Apr}, 
perturbation theory \cite{Anderson1970Dec}, 
Fermi liquid theory \cite{Nozieres1974Oct}
and the dynamical mean field theory  \cite{Georges1996Jan}, 
can tackle the Kondo problem and contribute to the qualitative understanding of the phenomenon at low energies,
all of them rely on approximating the electronic correlations to describe the energies near the Kondo energy scale.
The numerical renormalization group method (NRG) \cite{Wilson1975Oct, Bulla2008Apr},
considered to be the best at tackling the Kondo problem, can provide
quantitatively accurate description of the Kondo effect, but only
up to linear response studies near equilibrium \cite{Costi2010Jun}.
 
Notable theoretical attempts to describe the nonequilibrium transport through a Kondo impurity had employed 
nonequilibrium Greens function (NEGF)
\cite{VanRoermund2010Apr,Sierra2017Aug,Eckern2020Jan},
renormalized perturbation theory (RPT) \cite{Kaminski2000Sep}, 
generalized Fermi liquid theory
\cite{Oguri2001Sep,Mora2015Aug,Oguri2018Jan, Teratani2024Apr},
perturbative approaches \cite{Csonka2012May, Manaparambil2023Feb},
auxiliary master equation approach (AMEA) \cite{Dorda2016Dec,
Fugger2018Jan, Fugger2020Apr},
non-crossing approximation (NCA) \cite{Wingreen1994Apr, Hettler1998Sep}
and slave-boson mean field theory (SBMFT) \cite{Sierra2017Aug}. Though each method has
its own virtues and provides theoretical insights at various limits,
a complete picture of the whole nonequilibrium Kondo regime had remained elusive.
A hybrid method, incorporating both numerical renormalization group
and time-dependent density matrix renormalization group (tDMRG)
method based on a thermofield quench approach (NRG-tDMRG),
has achieved the feat of describing the nonequilibrium transport
through a Kondo-correlated system with exact treatment of correlations \cite{Schwarz2018Sep}.
Until now, this method has been employed to address the electronic transport under finite potential bias \cite{Schwarz2018Sep}
and spintronic transport in the presence of ferromagnetism in the leads \cite{Manaparambil2022Sep}.
In this work, we extend the NRG-tDMRG method to describe the nonequilibirum Kondo effect 
in the presence of finite temperature gradients.
In particular, we consider a quantum dot
symmetrically coupled to two metallic leads
held at different temperatures that can be tuned independently.
The choice of symmetric couplings to the leads allows for the Kondo correlations
to develop over both the leads, uncovering the influence of the
individual lead temperatures on the Kondo effect.
The dynamics of the electronic and heat  currents are calculated using NRG-tDMRG and 
their nonequilibrium steady state values are extracted using linear prediction across a finite time window. 
We characterize the Kondo regime as a function of the individual lead temperatures using 
the zero-bias conductance and the Seebeck coefficient of the system.
We find that transport in the presence of a nonlinear temperature gradient
can be qualitatively described by linear response results with an effective equilibrium temperature.
Our results demonstrate that the Kondo correlations persist
as a circle when depicted in the individual lead temperatures.

Our work provides the first quantitatively accurate results
for the thermoelectric transport coefficients of Kondo-correlated
quantum dot in far-from-equilibrium settings.
The paper is organized as follows: \Sec{Mod} describes the system Hamiltonian and the transport properties under study.
In \Sec{Res}, we discuss the results from NRG-tDMRG calculations.
We begin by discussing a noninteracting system in \Sec{RLM},
and then moving on to the interacting system described
by the single impurity Anderson model in \Sec{SIAM}.
The influence of temperature gradient on the zero-bias transport properties is discussed
in \Sec{epsd}. The differential conductance at zero-bias and for a finite potential bias
in the presence of different lead temperatures is discussed in \Sec{G}.
The thermoelectric current, Seebeck coefficient and heat transport properties are discussed in \Sec{S}.
Finally, the paper is summarized in \Sec{Summary}.

\section{\label{Sec:Mod}Model and Method}

\subsection{Hamiltonian}

\begin{figure}[h!]
 	\includegraphics[width=0.9\columnwidth]{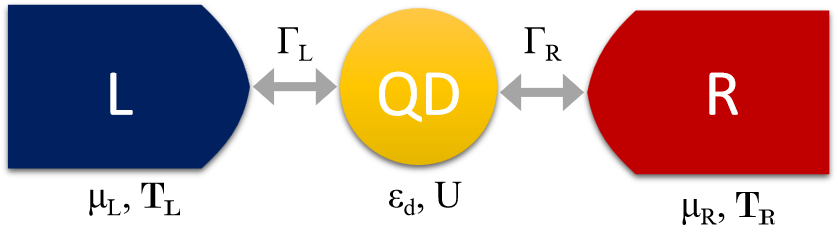}
 	\caption{
 	The schematic of a quantum dot with orbital level $\e_d$
 	and Coulomb repulsion $U$ coupled to 
 	the left ($\alpha=L$) and right ($\alpha=R$) metallic lead 
 	with hybridization function $\G_{\alpha}$.
 	Each lead is held at different temperature
 	$T_{\alpha}$ and chemical potential $\mu_{\alpha}=\pm V/2$.
    \label{fig:schem}}
\end{figure}

Our system consists of a quantum dot strongly coupled to two metallic leads.
The Hamiltonian of such a system can be described as
\be
H = H_{\rm imp} + H_{\rm lead} + H_{\rm tun},
\ee
where $H_{\rm imp}$  is the impurity part of the Hamiltonian described
by a single impurity Anderson model (SIAM)
with orbital energy $\e_d$ and Coulomb interaction $U$.
$H_{\rm imp}$ takes the form
\be
H_{\rm imp}=\e_d(n_{\up} + n_{\down}) + U\, n_{\up}n_{\down},
\ee
where $n_{\s}=d^{\dagger}_{\s} d_{\s}$ is the number operator,
with $d_\s$($d^{\dagger}_{\s}$) being
the annihilation (creation) operator for a dot electron with spin $\s$.
The leads are modeled as noninteracting particles
\be
H_{\rm lead} = \sum_{\alpha}H_{\alpha}=\sum_{\alpha k \s} \e_{\alpha k} c^{\dagger}_{\alpha k \s} c_{\alpha k \s},
\ee
with $c_{\alpha k \s}$ ($c^{\dagger}_{\alpha k \s}$) denoting
the annihilation (creation) operator for an electron
in the lead $\alpha$ with energy $\e_{\alpha k}$ and spin $\s$.
Finally, the tunneling Hamiltonian $H_{\rm tun}$
describes the coupling of the quantum dot to the leads
\be
H_{\rm tun}=\sum_{\alpha k\s}( v_{\alpha k} d^\dagger_\s c_{\alpha k \s} + {\rm H.c.}),
\ee
where $v_{\alpha k}$ is the tunneling matrix element
between the $k$th mode in the lead $\alpha$
and the quantum dot. The dot hybridizes with the leads
with the coupling strength given by, 
$\Gamma_\alpha = \pi\rho_\alpha |v_{\alpha k}|^2$,
where $\rho_\alpha$ denotes the density of states of the lead $\alpha$,
which is assumed to be flat ${\rho_\alpha\equiv 1/2D}$,
with $D$ being the band halfwidth
which is used as the unit of energy, hence $D=1$.
In the following, without loss of generality, we assume
that the system is symmetric $\Gamma_L = \Gamma_R = \Gamma$.
We set
\begin{eqnarray}
  \Gamma=0.001, \ \ U=12\,\Gamma, \ \ \varepsilon_d = -U/3,
\label{Eq:param}
\end{eqnarray}
unless specified otherwise.
The bias voltage $V$ is applied symmetrically as $\mu_L=-\mu_R=V/2$ and the left and right lead temperatures $T_L,T_R$ can be controlled independently.

To accurately take into account correlation effects
at truly nonequilibrium settings, 
we employ a hybrid NRG-tDMRG method in the matrix product state (MPS)
framework \cite{Schwarz2018Sep,Manaparambil2022Sep}. 
This method consists of a logarithmic-linear discretization scheme
of the conduction bands, a thermofield treatment, followed by a recombination of the leads modes,
and finally the time evolution by the second-order Trotter decomposition to reach the nonequilibrium steady-state.
The resolution of the method in the energy domain is conditioned by the number of intervals within the transport window.
The steady state values as $t \to \infty$ of heat and charge currents are found from linear prediction of finite time dynamics (cf. \App{prediction}).

Since the hybrid NRG-tDMRG approach involves a linear discretization within the transport window and then the time evolution in this discretized basis by tDMRG,
for realistic calculations in the case of a finite thermal bias,
this sets the limit on the difference
in the temperatures of the left and right leads to be around
two orders of magnitude.
More detailed description of the method
is presented in \App{NRG-tDMRG}.

\subsection{Transport coefficients}

The charge current $J_{\alpha\s}$ from the lead $\alpha$ to the quantum dot
in the spin channel $\s$ is given by
\bea
J_{\alpha \s} &=& e\, \langle \dot{N}_{\alpha \s}\rangle =-\frac{ie}{\hbar} \langle [N_{\alpha \s},H]\rangle \nonumber \\
&=&\frac{e}{\hbar} \sum_{k} {\rm Im} \, (v_{\alpha k}\, \langle d^{\dagger}_\sigma c_{\alpha k \s} \rangle ).
\eea
Here, $N_{\alpha \s} = \sum_{k} c^{\dagger}_{\alpha k \s} c_{\alpha k \s}$ is the occupation number in the lead $\alpha$.
Similarly, the energy current $J^{E}_{\alpha}$ from the lead $\alpha$
to the quantum dot can be described based on the lead Hamiltonian $H_\alpha$ as
\bea
J^E_{\alpha} &=& \langle \dot{H}_{\alpha} \rangle = -\frac{i}{\hbar} \langle[H_{\alpha},H]\rangle \nonumber \\
&=& \frac{1}{\hbar} \sum_{k\s} \e_{\alpha k} \, {\rm Im}\, (v_{\alpha k}\, \langle d^{\dagger}_\s c_{\alpha k\s}\rangle).
\eea
In the case of $V=0$, the energy current can be considered as the heat current $J^{Q}\equiv J^{E}$.
We note that since the symmetrized charge (heat) current
$J^{(Q)}(t)$ converges faster
than the current contributions from the individual leads, $J^{(Q)}_{\alpha \s}(t)$,
it is more efficient to find the steady-state value of the total current $J^{(Q)}(t)$,
\be
J^{(Q)} (t) = \sum_{\s}\frac{1}{2} [J^{(Q)}_{L\s} (t) - J^{(Q)}_{R\s} (t)].
\ee
More details about estimating $J^{(Q)} (t)$ and the steady state $J^{(Q)}$ can be found in \App{prediction}.

The differential electronic conductance $G$ 
and the electronic contribution to the heat conductance $\kappa$ 
are respectively defined as
\begin{eqnarray}
   G &=& \left(\frac{dJ}{dV}\right)_{T_L,\,T_R}, \notag \\ 
   \kappa &=& \left( \frac{J^{Q}}{\Delta T} \right)_{V}
\,.
\label{Eq:G_K}
\end{eqnarray}

The Seebeck coefficient $S$ estimates the potential $V$ required
to compensate for the induced thermoelectric current $J$
under a finite temperature gradient $\Delta T$ and it is defined as
\be
S = -\left(\frac{V}{\Delta T}\right)_{J=0}.
\label{Eq:S}
\ee

For the transport across an impurity coupled to metallic leads in the linear response regime,
these transport coefficients can be estimated as a function of the Onsager integrals,
$L_n = -\frac{1}{h} \int d\w (\w - \mu)^n \frac{\partial f}{\partial \w} \mathcal{T}(\w)$,
where $\mathcal{T}(\w)$ is the transmission coefficient of the impurity
and it is essentially equivalent to the equilibrium spectral function $A(\w)$ \cite{Haug}.
The linear response transport coefficients can thus take the form \cite{Costi2010Jun}
\bea
G_{0} & = &  e^2 \, L_0, \notag \\ 
S_{0} & = & -\frac{1}{eT} \frac{L_1}{L_0},\\
\kappa_{0} & =  &\frac{1}{T} \, \left( L_2 - \frac{L_1^2}{L_0} \right). \notag
\label{Eq:lin}
\eea

\section{\label{Sec:Res}Results and discussion}
In this section, we present and discuss the NRG-tDMRG results for the nonequilibrium
transport through a quantum dot in the presence of temperature gradients.
The details of the NRG-tDMRG calculations are described in \App{NRG-tDMRG}
where the method specific parameters are provided in \App{Evolution}.
First, the results for a noninteracting impurity under finite potential bias and 
temperature gradient are compared with exact results in \Sec{RLM}.
On the other hand, the nonequilibrium transport across an interacting impurity
in the presence of a finite temperature gradient is discussed in \Sec{SIAM}.
 
\subsection{\label{Sec:RLM} Noninteracting case: Resonant Level Model}

\begin{figure}[!ht]
 	\includegraphics[width=0.99\columnwidth]{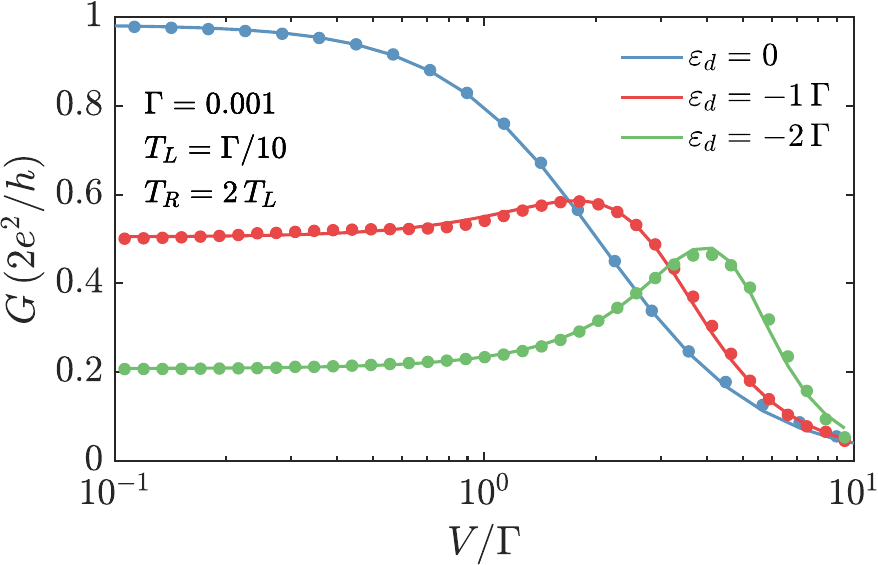}
 	\caption{
	The differential conductance
	for the	resonant level model ($U=0$)
	vs. potential bias $V$ 
    for fixed $\Delta T \sim T$
	(see model parameters to the left) and
    orbital energies $\e_d$ as indicated in the legend.
 	NRG-tDMRG data (dots) is compared to
	exact analytic curves for continuum (lines)
	as a consistency check.
	}
	
    \label{fig:G_RLM}
\end{figure}

As a benchmark for the nonequilibrium calculations,
we consider the noninteracting resonant level model (RLM),
i.e., essentially the Anderson model with $U=0$.
For this case, the current flowing through the system 
can be solved exactly \cite{Haug}
\be
J(V,\Delta T) = \frac{2e}{h} \int d\omega \;\mathcal{T}(\omega)\;[f_L(\omega)-f_R(\omega)],
\label{Eq:RLM}
\ee
where $\mathcal{T}(\w)$ denotes the transmission coefficient,
which can be related to the quantum dot spectral function $A(\w)$,
 $\mathcal{T}(\w) = \pi\Gamma A(\omega)$.
For the noninteracting quantum dot, the spectral function can be found
exactly through the equation of motion for the Green's function.
The transmission coefficient is then given by 
$ \mathcal{T}(\w)  = \G^2/(\G^2+(\w-\e_d)^2) $, where  $f_\alpha(\omega) = 1/\{1+\exp[( \omega - \mu_\alpha ) ]/T_\alpha \}$
is the Fermi-Dirac distribution function of lead $\alpha$ with $k_B\equiv 1$.
The differential conductance $G(V)$ calculated using NRG-tDMRG method for a
noninteracting  quantum dot
with different orbital level energies $\e_d$ in the presence of finite potential and
temperature gradients is shown in \Fig{fig:G_RLM}.
The differential conductance $G(V)$ has a peak around $V=2\, \e_d$, which is
attributed to the Lorentzian peak in $A(\w)$ located at $\w=\e_d$.
The shift in the differential conductance peak from the Lorentzian peak
originates from the symmetric nature of applied bias $\mu_L=-\mu_R=V/2$,
resulting in the transport window (TW) [$f_L(\w)-f_R(\w)$] inside the integral in \Eq{RLM} scanning the peak mainly
around $\w = 2 \, V$. It is important to note that the both temperatures
($T_L,T_R$) smear out the transport window and can thus only broaden the conductance
peak. The exact analytical calculations (lines) in \Fig{fig:G_RLM} agree perfectly
with the NRG-tDMRG data (dots), affirming that this technique can capture
the nonequilibrium transport primarily originating
from the nonlinear dependence of the lead Fermi distributions on $V$ and $T$.

\subsection{\label{Sec:SIAM} Interacting case: Single Impurity Anderson Model}

\begin{figure*}[t!]
 	\includegraphics[width=\textwidth]{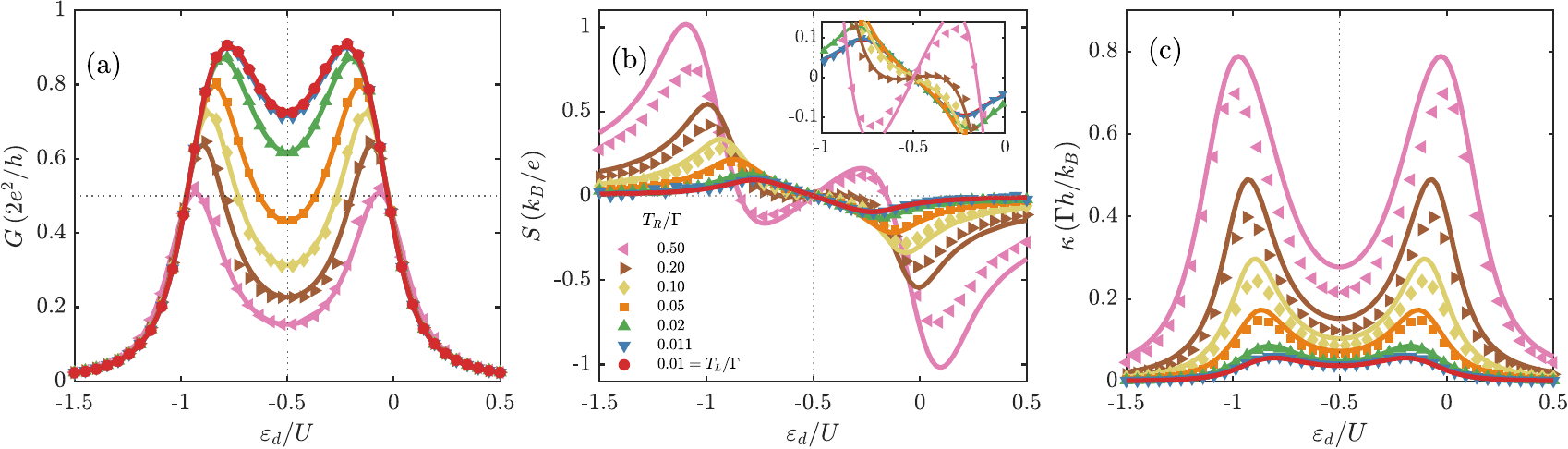}
 	\caption{(a) 
	    Differential conductance $G$, (b)
		Seebeck coefficient $S$, and (c)
		heat conductance $\kappa$
 		of an interacting quantum dot
		[SIAM using parameters \eqref{Eq:param} except for
		$\e_d$ which is varied here] 
	    vs. orbital level position $\e_d$
 		in the linear response regime with respect to the bias voltage.
 		The left lead is kept at temperature $T_L=0.01\,\Gamma$
		throughout,
 		while the right lead temperatures are specified with the legend in (b).
 		The colored symbols are the nonequilibrium data from the NRG-tDMRG calculations
 		and the solid lines present the equilibrium NRG data
 		with the same parameters but calculated for an effective global temperature
 		$T_{\rm eff}=\Trms$.
    Numerically, the determination of $S$ and $\kappa$
    require a finite temperature difference $\Delta T$.
    Hence, no red dots are
    shown for the case $T_L=T_R$ in (b) and (c). The limiting case
    $\Delta T \to 0$, however, is reflected in the small
    $\Delta T/T \sim 0.1$ data set (blue triangles),
    which already agrees well with the equilibrium NRG 
    data for $\Delta T=0$ (red line).
\label{fig:epsd}}
\end{figure*}

In the presence of finite $U$, the nonequilibrium transport across the quantum dot
becomes highly nontrivial and cannot be boiled down
to an analytical description without sufficient approximations \cite{Sierra2017Aug,Oguri2018Jan}. 
But, the linear response description of transport across an interacting quantum dot in equilibrium
can very well be calculated using the definitions in \Eq{lin}
once the spectral function $A(\w)$ is obtained.
The equilibrium spectral function $A(\w)$ of a SIAM with finite $U$
can be calculated using NRG with extreme precision, 
and thus it will be used as the benchmark for the calculation of linear response coefficients.
The NRG data discussed in this section have been calculated
using the QSpace tensor library for Matlab \cite{Weichselbaum2012Dec, Weichselbaum2020Jun, Weichselbaum2024May}
with discretization parameter $\Lambda=2$, iteration number $N=60$
and the maximum number of states kept $N_{\rm K}$ after each iteration as $2^{10}$.

\subsubsection{\label{Sec:epsd} Influence of finite temperature gradient}

We first introduce the finite temperature gradient across a SIAM
by keeping the left lead temperature at $T_L= 0.01 \G$ and
changing the right lead temperature from $T_R=0.01 \G$ to
$T_R=0.5 \G$. The electric current $J(V,T_L,T_R)$ and heat
current $J^{Q}(V,T_L,T_R)$ across the SIAM
using \eqref{Eq:param} 
is calculated for bias voltages close to linear response
$\Vlin \approx
0.005 \, \G$
using the NRG-tDMRG method.
Thus the differential conductance $G(T_L,T_R) \equiv G(V=0,T_L,T_R)$ can be estimated as
\begin{eqnarray}
    G(T_L,T_R) &=& \Bigl.\tfrac{1}{2\Vlin}\Bigl(
       J(\Vlin) - J (-\Vlin)
    \Bigr)\Bigr|_{T_L,T_R}.
\label{Eq:G:V0}
\end{eqnarray}
The choice of linear response bias voltage $\Vlin$ is such 
that any nonlinear behavior of $G(V)$ can be avoided,
allowing us to treat the estimated currents as linear in $V$.
Since the bias values $V=\pm \Vlin$ are effectively in the linear response regime,
the charge (heat) current  $J^{(Q)} $ at zero bias
can be calculated according to the linear response expansion as,
\be
   J^{(Q)} (T_L,T_R) = \tfrac{1}{2} \Bigl.\Bigl(
      J^{(Q)}(\Vlin) + J^{(Q)} (-\Vlin)
      \Bigr)\Bigr|_{T_L,T_R}.
\ee
The electronic contribution to the heat conductance according to \Eq{G_K}
can thus be $ \kappa (T_L,T_R) = J^Q (T_L,T_R) /(T_R-T_L)$.
The information about $J(T_L,T_R)$ and $G(T_L,T_R)$ at $V=0$
is sufficient to calculate the Seebeck coefficient $S$ for the respective parameters. 
Moreover, the linear response in $V$ allows the current for small voltages
to be expressed as $J(V)=J(0) + V\, G$ for constant $T_L$ and $T_R$.
Thus the Seebeck coefficient $S$ from its definition in \Eq{S} can be estimated as,
\be
S (T_L,T_R) = -\frac{1}{T_R-T_L}\frac{J(T_L,T_R)_{V=0}}{G(T_L,T_R)_{V=0}}.
\label{Eq:S_num}
\ee

The transport coefficients for a quantum dot in the presence of finite temperature
gradient calculated using NRG-tDMRG are shown in \Fig{fig:epsd}. The differential
conductance $G$ seen in \Fig{fig:epsd}(a) shows the evolution of the zero-bias
conductance peak as a function of the orbital energy $\e_d$. The red dots in
\Fig{fig:epsd}(a) display the NRG-tDMRG data for $T_L=T_R=0.01\,\G$,
which match exactly with the equilibrium NRG data (red curve) for $G_{0}$
computed with a global temperature $T=0.01\G$.
The large conductance inside the local moment regime, $-U\lesssim \e_d\lesssim 0$,
is a characteristic feature of the Kondo resonance
and the thermal fluctuations from the leads with temperature $T=0.01\,\G$
limit the conductance from reaching the unitary value of  $G_0=2e^2/h$. 

The Kondo temperature $\TK$ in the local moment regime is
analytically given by
the improved Haldane formula from Fermi liquid theory \cite{Mora2015Aug}
\be
\TK = \sqrt{\frac{\Gamma\,U}{2}}\, {\rm exp} \left[\frac{\pi\,\e_d\,(\e_d+U)}{2\,\Gamma\,U} + \frac{\pi\,\Gamma}{2\,U}\right].
\label{Eq:Haldane}
\ee

Since the Kondo temperature represents a crossover scale,
it is only defined up to a prefactor of order one.
Hence, alternatively from a data or experimental point of view,
the Kondo temperature can be estimated by the temperature
at which the zero-bias conductance drops by half.
Below, we will refer to this as $T_{K'}$, where
based on our data for the parameters in \Eq{param},
$T_{\rm K'} \simeq 1.05\, \TK$ [cf.~\Fig{fig:G}(c)
and caption]. 

For the SIAM parameters in \Eq{param},
we have $\TK = 0.042 \, \Gamma$ [as compared to
the lowest value at $\e_d = -U/2$, $\TK = 0.025 \, \G$].
Thus, in the local moment regime, the $G(\e_d)$ curves in \Fig{fig:epsd}(a)
show minima at $\e_d=-U/2$ corresponding to the lowest $\TK$.
We proceed to heat up the right lead ($T_R$),
as specified in the legends of \Fig{fig:epsd}.
With increasing $T_R$,
the differential conductance in the local moment regime
decreases as the Kondo resonance dies off with increasing thermal fluctuations from the hotter lead. 
The equilibrium NRG cannot account for different lead temperatures,
but one can still define an effective global temperature $T_{\rm eff}$
at equilibrium as the root mean square value of the left and right lead temperatures
\be
 T_{\rm eff} = \Trms=\sqrt{\tfrac{1}{2} (T_L^2+T_R^2)}.
 \label{Eq:rms}
\ee
The significance of the root mean square value will be discussed in the next section, \Sec{G}.
For the sake of the discussion here, it is sufficient to note that $\Trms \to T$ when $T_R \to T_L$.

In \Fig{fig:epsd} we show that a striking agreement exists between
the nonequilibrium NRG-tDMRG results at finite thermal bias (colored symbols)
and the equilibrium NRG results with an effective global temperature $\Trms$
defined as the root mean square value of the lead temperatures.
Implying that the dependence on the individual lead temperatures
mimics the dependence of equilibrium Kondo resonance width with a global temperature $\Trms$.
This is consistent with the low temperature
limit from the perturbation theory and slave-boson mean field theory
results of Ref.~\cite {Sierra2017Aug}. 
Moreover, the NRG-tDMRG results are valid for higher temperatures
due to the exact treatment of correlations.
It is also interesting to note that this effective $\Trms$ equivalence extends even into the mixed
valence and empty/filled orbital regimes ($\e_d\lesssim -U$, $\e_d \gtrsim 0$).
Furthermore, the experimental works for the thermoelectrics
in the Kondo regime show a good agreement with our results.
Figure~2 of Ref.~\cite{Svilans2018Nov}, showing
the differential conductance and thermocurrent with 
$\Delta T/T \approx 0.3$ for different $T$ near the Kondo regime, 
behaves in a very similar way to the results presented here.
On the other hand, the experimental data for the Seebeck coefficient
shown in Fig. 4 of Ref.~\cite{Dutta2019Jan} were related to the linear response NRG results.
We note that in this case, the corresponding temperature gradients,
though not precisely determined due to the experimental conditions,
reached $\Delta T/ T \approx 2/3$ which is well beyond
linear response theory.
The qualitative agreement obtained with
linear-response NRG, nevertheless,
we attribute to the $\Trms$ equivalence discussed in this
paper. A more quantitative agreement can be obtained using $T=\Trms$
in the NRG calculations, provided that the temperatures of the individual leads are known.
Of course, deep in nonequilibrium, i.e.,
much beyond linear response, 
one needs to resort to out-of-equilibrium approaches
such as NRG-tDMRG.

The linear response Seebeck coefficient $S_{0}$ of a quantum dot as a function
of the global temperature has been shown to change sign with the onset 
of the Kondo correlations \cite{Costi2010Jun,Sierra2017Aug,Svilans2018Nov,Dutta2019Jan}.
On the other hand, the nonlinear temperature gradient dependence of $S$
in the Kondo regime is largely unknown. Here, with our NRG-tDMRG method,
we are able to provide first accurate data on it,
which are presented in \Fig{fig:epsd}(b).
The red curve represents the equilibrium case where $\Trms=T_L=T_R=0.01\,\G$.
Note that the calculation of $S$ from NRG-tDMRG requires
a finite temperature gradient according to \Eq{S_num}
and thus nonequilibrium data is absent for the $T_L=T_R$ case.
The representative linear response results from NRG-tDMRG
are presented in the case of $T_R=0.011\,\G$ (blue traingles)
and agree well with the equilibrium results from NRG.
The Seebeck coefficient remains antisymmetric across the particle-hole symmetry point
$\e_d=-U/2$ and has a non-zero value in the local-moment regime,
as expected for finite temperatures below $\TK$.

When the right lead temperature is increased, i.e. with a finite thermal bias,
the Seebeck coefficient becomes reduced and starts to change sign
in the local moment regime around $T_R = 0.2\, \Gamma$, indicating the destruction of the Kondo resonance.
Interestingly and quite unexpectedly, the comparison to the equilibrium NRG results with a global temperature
$T_{\rm rms}$ gives a reasonably good agreement in the local moment regime.
The sign change in equilibrium $S_{0} (T)$ 
occurs at higher temperatures than $\TK$, which is also reflected in our finite $\Delta T$ results.
However, outside the local moment regime,
where the Kondo correlations do not emerge,
the Seebeck coefficient increases in magnitude (no sign changes)
with the increase in $T_R$ and, correspondingly, with $\Trms$.
In this regime, the effective linear response results show
growing deviations from the nonequilibrium results with increasing temperature gradient.

Finally, for the sake of completeness,
we examine the heat conductance $\kappa$ as a function of $\e_d$ in \Fig{fig:epsd}(c).
The heat conductance is dominated by the contribution associated with charge fluctuations,
which are most active at resonances.
As can be seen, $\kappa$ generally has two peaks
corresponding to the proximity of the resonant levels to the Fermi energy
at $\e_d \approx 0$ and $\e_d \approx -U$.
With a finite thermal bias, $\kappa$ shows deviations from the linear response $\Trms$ calculations
that increase with raising the temperature gradient.

\subsubsection{\label{Sec:G}The Kondo circle}
   
\begin{figure*}[ht!]
	\centering
 	\includegraphics[width=0.99\textwidth]{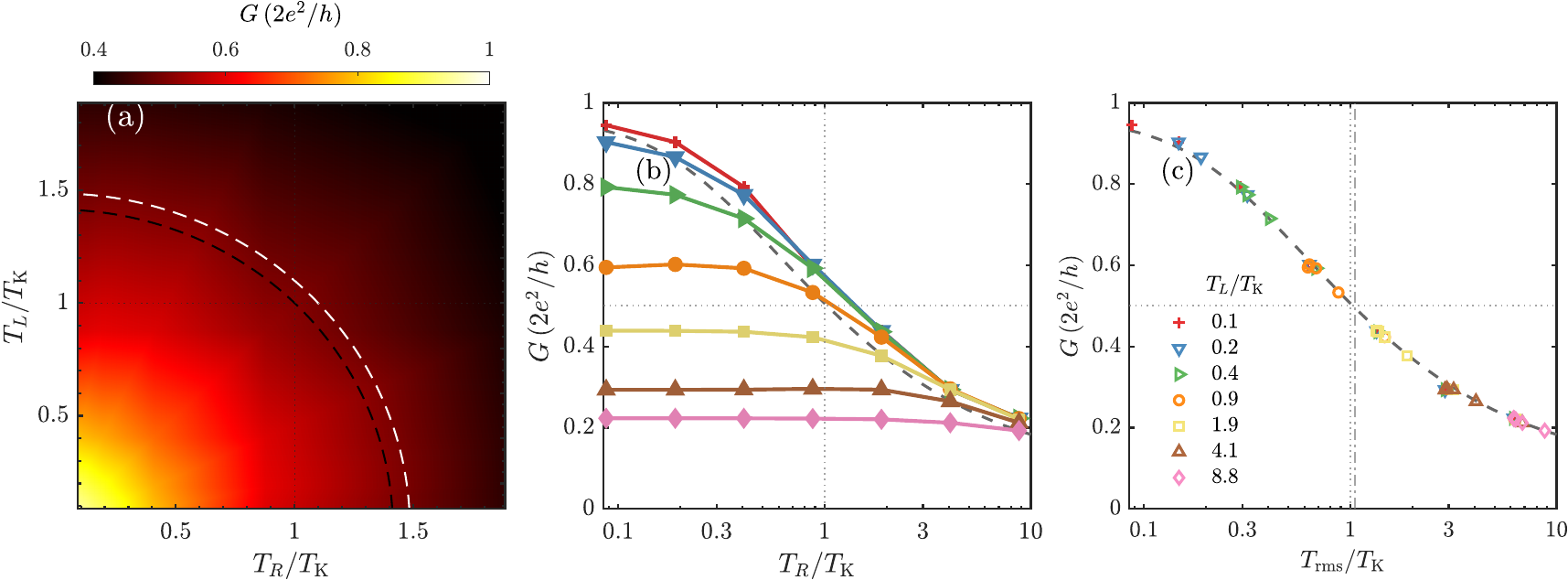}
 	\caption{
 	(a) The differential conductance $G$ through the quantum dot
 	with orbital energy $\e_d=-U/3$ 
 	as in \eqref{Eq:param}
    as a function of the left and right lead temperatures,
 	$T_L$ and $T_R$, in the linear response regime
	$V\to 0$.
 	The black dashed curve shows a circle of radius $\sqrt{2}\, T_K$ 
 	corresponding to  $\Trms = \TK$ [cf. \Eq{Haldane}],
 	while the white dashed line shows
 	$\Trms = T_{\rm K'}$ estimated as the half-width of the zero-bias conductance peak from the NRG
 	data versus effective temperature.
 	The colored symbols in (b) present 
	horizontal cross-sections
 	of (a) for different values of
	$T_L$, as shown with panel (c) vs.	$T_R$ on a
	logarithmic scale.
 	For comparison, the black dashed line
	displays the linear response NRG results of $G$ 
	vs. $T_R=T_L \equiv T$.
 	Panel (c): Data in (b) replotted against
 	the effective global temperature $\Trms$ in \Eq{rms}. 
	This is again contrasted with 
	the equilibrium NRG data
	(black dashed line) where the vertical dash-dotted
	 line denotes the half-width of equilibrium conductance
	$T_{\rm K'} \simeq 1.05 \, \TK$.
 	\label{fig:G}}
 	
 	\includegraphics[width=0.99\textwidth]{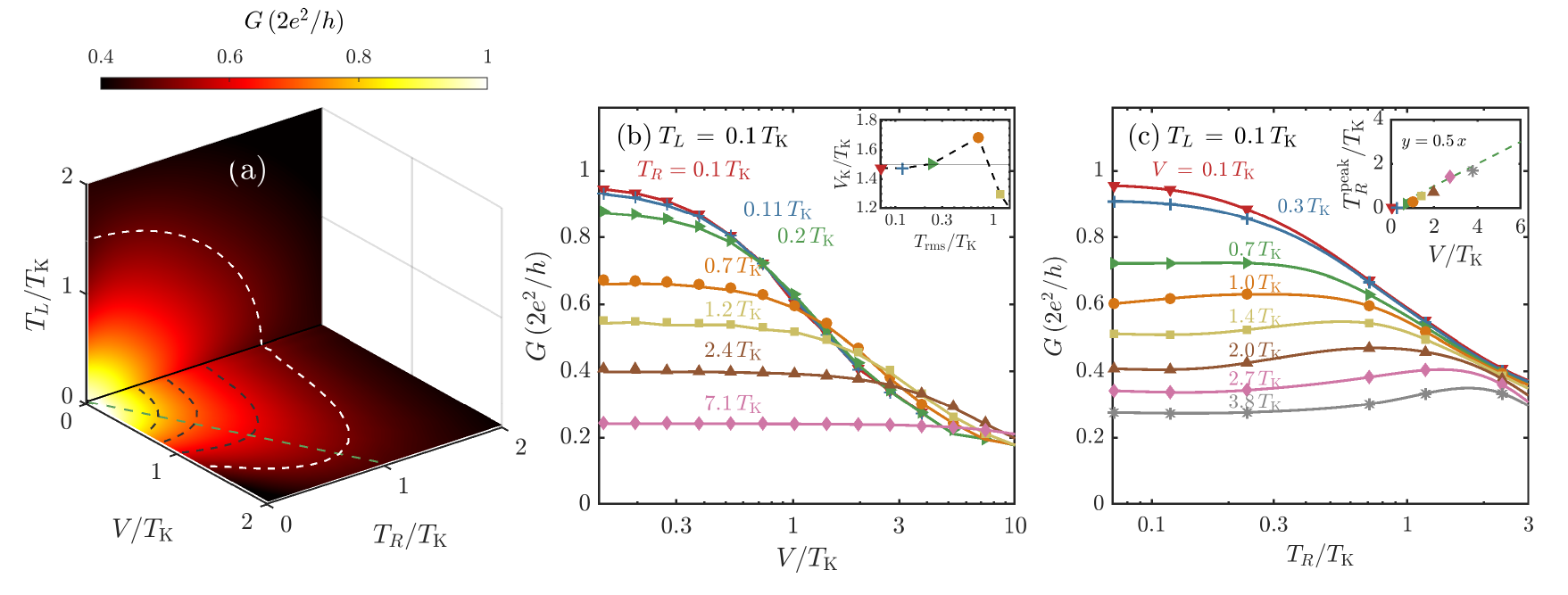}
 	\caption{
	(a) The differential conductance $G$ through a quantum dot
    [SIAM using \eqref{Eq:param}]
    vs. $T_L$, $T_R$ and a finite potential bias $V$. The data in the vertical plane is the same
     as in \Fig{fig:G} (a), the horizontal plane is calculated with $T_L=0.1\, \TK$ 
     and for different $T_R$ as specified in the legends of panel (b). The black [white]
      dashed lines on the horizontal plane show the contours of constant conductance
      $G=(0.8, 0.7, 0.6)$ [$G=0.5$].
     The green dashed line indicates $T_R = V/2$,
     cf. inset to panel (c).
     Panel (b) shows the cross-sections (symbols) of the horizontal plane
     in panel (a) for a fixed right lead temperature as indicated by the colored labels. 
     The solid lines show the corresponding $G(V)$ calculations for an effective global 
     lead temperature $\Trms=T_L=T_R$. The inset in panel (b)
     tracks $\VK$, the Kondo scale in the applied bias, defined as $G(\VK)= 0.5$.
     Panel (c) shows the differential conductance $G(T_R)$ from the horizontal plane in panel
     (a) for a finite potential difference $V$, as indicated by the colored labels. 
     Lines represent spline interpolations of the semilog-x data
     used to estimate the peak position $T_R^{\rm peak}$.
     Inset shows $T_R^{\rm peak}$ vs. $V$, which approximately
     follows $T_R^{\rm peak} = V/2$ (green sdashed line).
     \label{fig:GV}}
\end{figure*}

In this section, we discuss how the Kondo effect
depends on the individual lead temperatures.
For this, we choose the orbital level $\e_d=-U/3$,
for the system to be in the local moment regime, but far enough from the particle-hole
symmetry point to develop sufficient thermopower $S$.

Figure~\ref{fig:G}(a) presents the zero-bias differential conductance $G$
as a function of the independent left and right lead temperatures.
The conductance $G$ has its maximum as $T_L,T_R \to 0$ and decays radially in the $T_L-T_R$
plane. In particular, we focus on the temperatures in the scaling regime, i.e. around $T=\TK$,
where the conductance $G_0(T)$ is known to exhibit universal behavior. 
The black [white] dashed curves denote circles of radii $\sqrt{2} \, \TK$,
where $\TK$ is estimated from \Eq{Haldane}
[$\sqrt{2} \, T_{\rm K'}$, where $T_{\rm K'}$ is estimated as the half-width
of the linear-response conductance $G_{0}(\Trms)$].
Though the Kondo temperature $T_{\rm K'} \approx 1.05 \,T_{\rm K}$
from the numerical NRG data provides a more accurate
approximation of the Kondo energy scale than the analytical formula,
for the sake of generality and ease of estimation, 
we will stick to $\TK$ as the definition of Kondo temperature in this paper.
Therefore, the half-width of the conductance peak lying
on the $T_{\rm K'}$ circle is an immediate consequence
from the definition of $\Trms$ and its correspondence
to the nonlinear $\Delta T$ in the local moment regime [cf. \Fig{fig:epsd}(a)].
The horizontal cross-sections in panel (b) show how the conductance
decays as a function of the right lead temperature $T_R$,
where the temperature on the left lead $T_L$
determines the peak value of the conductance curve.
The $G(T_R)$ curve lies below the linear response $G_{0}
(T_R=T_L=T)$ curve for $T_R<T_L$
and coincides with the linear response results at $T_L=T_R$
to remain above the linear response data for $T_R>T_L$.
Due to the left-right symmetry in the system,
the previous arguments hold true even if one swaps $T_L$ and $T_R$.
The conductance data $G(T_L,T_R)$ is plotted against
the rescaled $\Trms$ temperature in the panel (c). 
The rescaled data lies perfectly on top of the linear response $G_{0} (\Trms)$ curve.
This is a useful result, especially for the experimental exploration of the Kondo regime.
In experiments, where one do not reach the truly linear response regime
\cite{Svilans2018Nov, Dutta2019Jan, Gehring2021Apr, Hsu2022Apr},
$T_{\rm rms}$ can provide reliable theoretical estimations from equilibrium NRG calculations
to accurately identify the parameter space of the Kondo regime in $T_L$ and $T_R$ separately.

In general, the zero-bias conductance peak along with the Kondo
resonance is known to get smeared with increasing thermal
fluctuations \cite{Cronenwett1998Jul,Goldhaber-Gordon1998Dec}. The influence
of the individual lead temperatures on the whole $G(V)$ curve
beyond linear response bias voltage regime is less trivial and is shown in
\Fig{fig:GV}(a). The lower plane in \Fig{fig:GV}(a) presents
the $G(V)$ calculations for a cold left lead temperature
$T_L \approx 0.1\,\TK$ and with increasing the right lead temperatures
$T_R >T_L$. For small temperatures $T_R \ll \TK$, the
conductance peak remains sharp in the finite $V$ regime but with
an increase in $T_R$ around $T_R \approx 0.2\, \TK$ the Kondo peak
starts to get smeared out in $V$. This behavior is clearly seen
in $G(V)$ curves for different $T_R$ presented in
\Fig{fig:GV}(b), where the increase in $T_R$ suppresses the
conductance at zero bias and smears the zero-bias conductance
peak further into the finite $V$ regime. 

Furthermore, we observe in our simulations that any
configuration of the lead temperatures $G(V)_{T_L,T_R}$ 
can be approximated by a $G(V)_{\Trms,\Trms}$ curve with global temperature $\Trms$
[cf. solid lines in \Fig{fig:GV}(b)].
The Kondo energy scale in the applied bias $\VK$, defined
as the bias at which the conductance drops to one-half
$G(\VK)=1/2$, is a characteristic energy scale of the
nonequilibrium Kondo effect and behaves differently from $\TK$. 
The inset in \Fig{fig:GV}(b) shows the dependence of $V_{\rm K}$ on $\Trms$.
At low temperatures $T \ll \TK$, we recover the Fermi liquid theory
prediction for the Kondo energy scales $\VK/\TK \approx 3/2$ \cite{Mora2015Aug,
Schwarz2018Sep,Filippone2018Aug, Manaparambil2022Sep}.
It can be seen that $\VK$ increases with $\Trms$, corresponding to the smearing of the Kondo resonance
with thermal fluctuations up to $\Trms \approx \TK$.
Beyond which the Kondo resonance is considerably destroyed by the thermal fluctuations,
such that $G(V)$ fails to attain the definition of $\VK$ for temperatures around 
$\Trms \approx 1.3 \, \TK$ [cf. \Fig{fig:GV}(b)].

Figure \ref{fig:GV}(c) shows the influence of the right lead temperature $T_R$ on
the differential conductance $G(T_R)_{T_L,V}$ with a constant $T_L$ and
finite potential bias $V$.
For very small potential biases $V\ll\TK$, the differential conductance $G$ 
monotonously decreases with increasing $T_R$, closely resembling the true zero-bias conductance curve in 
\Fig{fig:G}(b). In the case of a large potential bias $V\gtrsim \TK$, 
the $G(T_R)$ curves show maxima roughly located at a finite right lead temperature $T_R^{\rm peak} \approx V/2$
[cf. inset of \Fig{fig:GV}(c)].
This nonmonotonous behavior of $G(T_R)_{T_L,V}$ for $V>\TK$ can be attributed to the 
splitting of the Kondo resonance in the presence of large potential biases.
Due to the bias configuration in our system, $\mu_{L/R} = \pm V/2$,
the peaks of the split-Kondo resonance will be located at the
respective lead potentials $\mu_{L/R}$ for $V\gtrsim\TK$,
resulting in the additional feature in $G(T_R)$ around $T_R=V/2$.

\subsubsection{\label{Sec:S}Thermoelectrics of the Kondo circle}

\begin{figure*}[ht!]
 	\includegraphics[width=\textwidth]{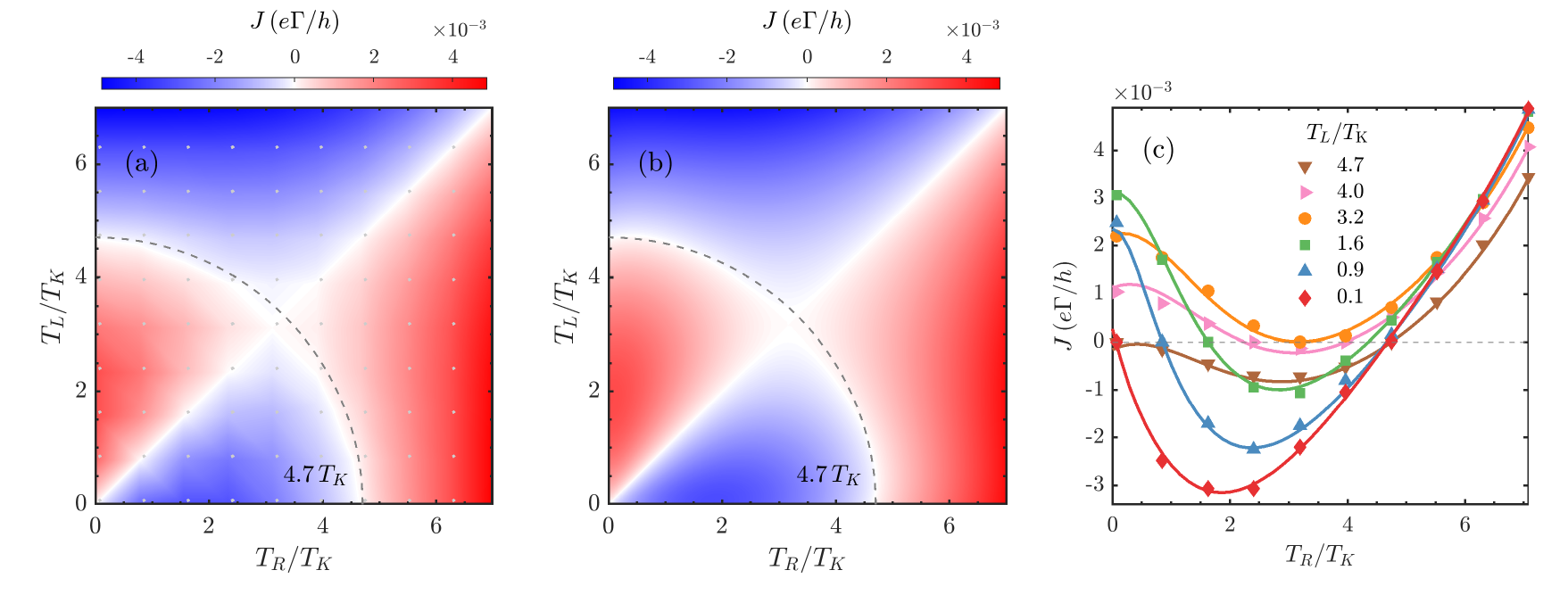}
 	\caption{
	(a) Thermoelectric current $J$ through a quantum dot
    [SIAM using \eqref{Eq:param}]
    vs. $T_L$ and $T_R$ at $V=0$ (computed as the 
    average current for $V=\pm 10^{-2} \TK$).
    The data points located at the white dots are interpolated
    by the smooth color shading (cf. color bar).
    Panel (b) same as panel (a), but showing the polynomial fit
    of its data points based on \Eq{polyfit} instead.
    Panel (c) shows horizontal cuts
    of the polynomial fit (lines) in panel (b) with their corresponding data points (symbols) from panel (a).
    \label{fig:J}}
 	\includegraphics[width=\textwidth]{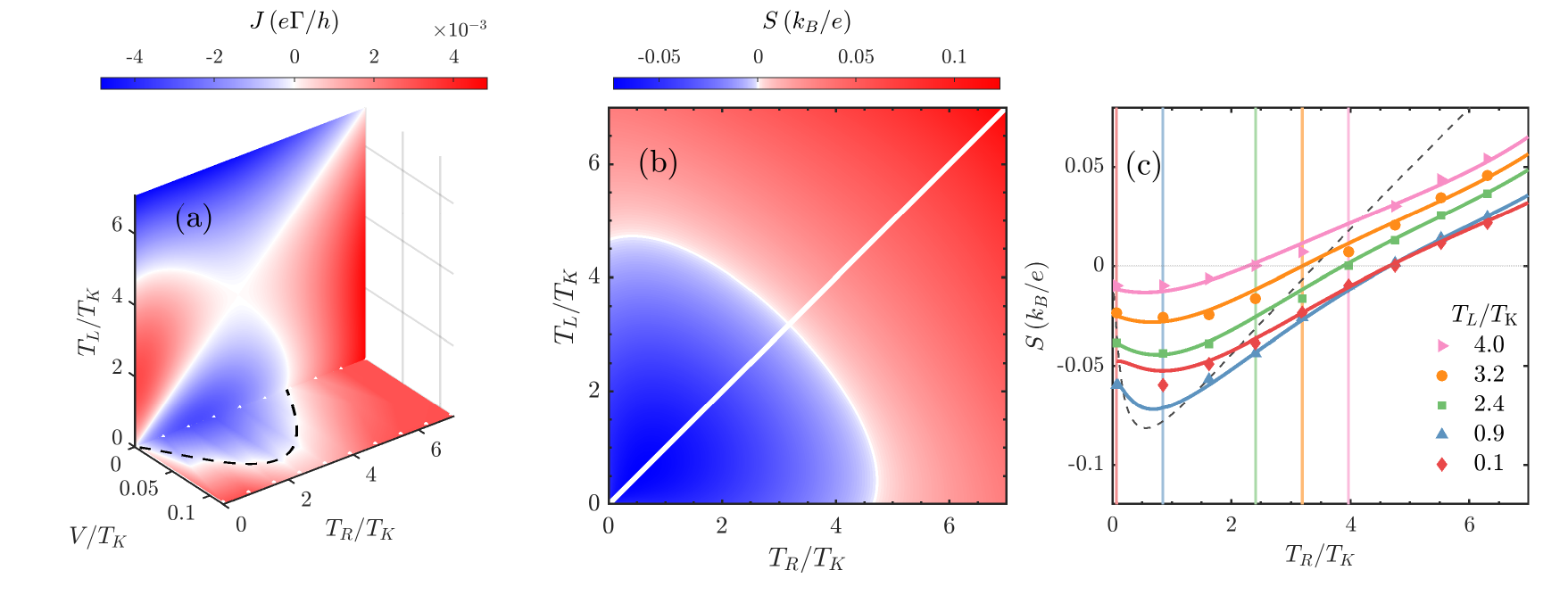}
 	\caption{
	(a) Thermoelectric current $J$ through a quantum dot
    [SIAM using \eqref{Eq:param}]
    vs. $T_L$, $T_R$ and for very small potential biases $V$.
    The black dashed curve shows the voltage $V$ required to compensate the thermocurrent according to
    a linear response expansion with $G_{0}$. The grid of NRG-tDMRG data in $V$ is
    represented as white dots on the $V,T_R$ plane. (b) The Seebeck coefficient 
    $S =-V/\Delta T = -J/(G\, \Delta T)$
    (assuming linear response in $G$) for the same parameters as in panel
    (a). Panel (c) shows the horizontal cross-sections of the panel (b) comparing $S$
    calculated from the fit [cf. \Eq{polyfit}] (solid line)
    and $S$ estimated from $J(T_L,T_R)_{V=0}$ from NRG-tDMRG (colored symbols).
    For comparison, the black dashed curve shows the linear response 
    NRG calculations for $S_0$ with $T_L=T_R$.
    The colored vertical lines denote the corresponding left lead temperature
    $T_L$ for each $S(T_R)$ curve.
    \label{fig:S}}
\end{figure*}

Instead of diving directly into the Seebeck coefficient,
we first look at the thermoelectric current driven by the finite thermal bias in \Fig{fig:J}.
The panel (a) shows the NRG-tDMRG results for the charge current
as a function of both the left and right lead temperatures.
The current $J(T_L,T_R)_{V=0}$ is antisymmetric across the $T_L=T_R$ line,
as the temperature gradient changes sign across this line.
In addition, there exists another sign change roughly as a circle in the $T_L,T_R$
plane corresponding to the onset of Kondo correlations.
The current at zero bias, computed as
$J(T_L,T_R)_{V=0}  =\left.\tfrac{1}{2}\left(
       J(\Vlin) + J (-\Vlin)
    \right)\right|_{T_L,T_R}$
from the data for small $\pm V_0$ [cf. \Eq{G:V0}],
can be fitted by the polynomial expression up to order $n$ as in 
\begin{eqnarray}
    J(T_L,T_R)_{V=0} &=& \Gamma \, \tfrac{T_L-T_R}{\Trms}\  
    p_n(x\equiv \tfrac{T_L}{\TK}, y\equiv \tfrac{T_R}{\TK}),
\label{Eq:polyfit}
\end{eqnarray}
where
\begin{eqnarray}
  p_n(x,y) &=& \sum_{k=1}^n \sum_{i=0}^{k} a_{k,i} \, x^i \, y^{k-i},
\label{Eq:polyfit:2} \\
  a_{k,i} &=& a_{k,k-i}
\ . \label{Eq:polyfit:3}
\end{eqnarray}
Having $V=0$, the current needs to be antisymmetric
under inversion $T_L \leftrightarrow T_R$. This is taken
care of by the leading factor $T_L-T_R$ on the RHS.
The remaining polynomial $p_n(\tfrac{T_L}{\TK},\tfrac{T_R}{\TK})$ thus must be
symmetric under inversion. This constrains
the polynomial terms to \Eq{polyfit:3}.
The denominator $\Trms$ keeps the prefactor in check for large $\Delta T$.
i.e., the ratio $\tfrac{T_L-T_R}{\Trms} \to \sqrt{2}$ as $T_R \to
 \infty$. Thus providing a much more consistent weights for the data points with large 
$\Delta T$ used in the variational fitting.
We note that a clean polynomial fit of the form $(T_L-T_R)\,p_n(\tfrac{T_L}{\TK}
,\tfrac{T_R}{\TK})$ can still provide an acceptable fit for the current,
but including the denominator $\Trms$ considerably improves the fit at low $T$.
At first glance, \Eq{polyfit} only seems to account for the first order in
$\Delta T$. But, the first order polynomial terms $T_L$,$T_R$ together with the $T_L-T_R$
prefactor makes up the $(T_L-T_R)^2 \equiv \Delta T^2$ dependence, 
the polynomial terms $T_L^2,T_R^2$ and $T_L\,T_R$ have encoded in it the
information of the $\Delta T^3$ dependence, and accordingly for the higher order dependences
in $\Delta T$. Thus the polynomial fit contains, but is not limited to, the perturbative 
expansion of $J$ on $\Delta T$.

The polynomial coefficients are determined
by minimizing the cost function
\begin{eqnarray}
   C &=& \sum_i 
   \Bigl| J(T_L,T_R)\bigr|_i - \Gamma \tfrac{(T_L-T_R)}{\Trms}\bigr|_i p_n(x_i,y_i)\Bigr|^2,
\end{eqnarray}
where the sum runs over all data points $i$ with $T_L\neq T_R$.
The quality of the fit is then estimated by the error measure 
$\delta_{\rm fit} = \sqrt{{\rm min}(C)}$.
The fit in \Fig{fig:J} used $n=4$ with coefficients
\bea
   (a_{10}\phantom{,a_{41},a_{42}}) & = & (2.7874) \notag \\
   (a_{20},a_{21}\phantom{,a_{42}}) & = & (-1.0856, -0.9690) \notag \\
   (a_{30},a_{31}\phantom{,a_{42}}) & = & ( 0.1363, 0.1418) \notag \\
   (a_{40},a_{41},         a_{42} ) & = & (-0.0068, -0.0091, -0.0060) 
   \label{Eq:coeff}
\eea
\begin{table}[h!]
\begin{center}
\begin{tabularx}{0.8\columnwidth}{| >{\centering\arraybackslash}X | >{\centering\arraybackslash}X|}
\hline
\vspace{0.0em}
$n$ & 
\vspace{0.0em}
$\delta_{\rm fit}/\bar{J}$ \\[0.2em]
\hline
1 &  $0.9690$\\[0.1em]
2 &  $0.0051$\\[0.1em]
3 &  $0.0010$\\[0.1em]
4 &  $0.0002$\\[0.1em]
\hline
\end{tabularx}
\end{center}

\caption{The degree $n$ of the polynomial
   used for the fit and corresponding error $\delta_{\rm fit}$ relative to $\bar{J}$ 
   the largest value of thermoelectric current inside the Kondo circle.
}
\label{err_table}
\end{table}

The thermoelectric current from the polynomial fit \Eq{polyfit} is shown in Figs.~\ref{fig:J}(b,c).
The polynomial fit accurately recovers the regions of sign change in \Fig{fig:J}(a).
The error measure of the fits presented in Table~\ref{err_table} shows that 
increasing order of the polynomial improves the fit quality. 
The fit converges at higher orders of the polynomial, indicated by the
decreasing magnitude of the polynomial coefficients for the higher order terms [cf. \Eq{coeff}].

The estimation of $S$ from \Eq{S_num} relies on the induced thermocurrent being small enough
to be compensated by a linear response bias $V$. 
In \Fig{fig:S}(a) we show the extension of the density plot in \Fig{fig:J}(a)
towards the third dimension in the bias voltage $V$.
The lower plane in $V, T_R$ is calculated for $T_L \approx 0.1\,\TK$ [brown curve in \Fig{fig:J}(c)],
which contains the largest value of thermocurrent data in the Kondo regime. 
The points of zero current in the lower plane show that a bias voltage  $V<\Vlin$ is
sufficient to compensate for the induced thermocurrent. 
The zero current (white) in the interpolated colormap from the NRG-tDMRG data for finite
$V=\pm \Vlin$ coincides with the bias estimated from the linear response
expansion (black dashed curve) of the current with the linear response
conductance $G_{0} (\Trms)$, 
further corroborating the choice of the linear response $\Vlin$.
Thus, \Fig{fig:S}(b) depicts the Seebeck coefficient $S$ 
estimated for the full scaling regime in $T_L,T_R$ plane.
From the sign changes of the thermoelectric current $J(T_L,T_R)$ in \Fig{fig:J},
only the sign change corresponding to the Kondo correlations survive for $S(T_L,T_R)$.
This region of the sign change in the Seebeck coefficient now fully represents 
the temperature regime in which the Kondo correlations survive.
The Kondo regime is roughly a circle in the $T_L,T_R$ plane,
slightly squeezed in the $T_L=T_R$ direction.
It is important to note that the radius of the Kondo regime in the $T_L,T_R$ plane
determined by the points of sign change in $S$
does not show any universal scaling with respect to $\TK$.
The equilibrium NRG studies of $S_0$ have already demonstrated that the temperature 
at which $S_0(T)$ shows the maximum negative value in the Kondo regime scales with the 
Kondo temperature $\TK$.
But the temperature at which $S_0$ changes sign, 
denoting the onset of Kondo correlations, does not 
exhibit such scaling with respect to $\TK$ \cite{Costi2010Jun}.

The quantitative behavior of $S(T_L,T_R)_{V=0}$ is shown in \Fig{fig:S}(c).
The Seebeck coefficient $S$ estimated from the NRG-tDMRG calculations (colored symbols)
of the thermoelectric current $J(T_L,T_R)_{V=0}$ in \Fig{fig:J}
and $S$ estimated from the polynomial fit for the thermoelectric current (solid lines)
as in \Eq{polyfit} with a constant $T_L$ are plotted as a function of $T_R$. 
Near the equilibrium temperature $T_R \to T_L$, the NRG-tDMRG results approach the 
linear response NRG estimations of $S_0$.
We note that, since $T_L=T_R$ induces no thermoelectric current, 
the extraction of the linear response $S_0$ using NRG-tDMRG from the chosen $T_L,T_R$ 
grid of discrete datapoints is not possible [cf. \Eq{S_num}], and hence no datapoints
from NRG-tDMRG are shown for the case of $T_L=T_R$ in \Fig{fig:S}(c).
The polynomial fit for the thermoelectric current from \Eq{polyfit} is 
unrestrained and can provide an 
approximation of the linear response $S_0$ for $T_L\to T_R$.  $S(T_R\to T_L)$
estimated from the fit shows slight quantitative difference from the true linear
response $S_0$ obtained from NRG, presumably stemming from the absence of very small $\Delta T$ in the
data used for fitting.
In general, for a constant $T_L$ in the Kondo regime, $S(T_R)$ starts from a 	 
negative value for $T_R \to 0$ and shows a minima at temperature $T_R$ of the order of
$\TK$. With further increase in the temperature, $S(T_R)$ grows gradually until changing
its sign denoting the total destruction of the Kondo resonance.
 
\begin{figure}[t!]
 	\includegraphics[width=0.99\columnwidth]{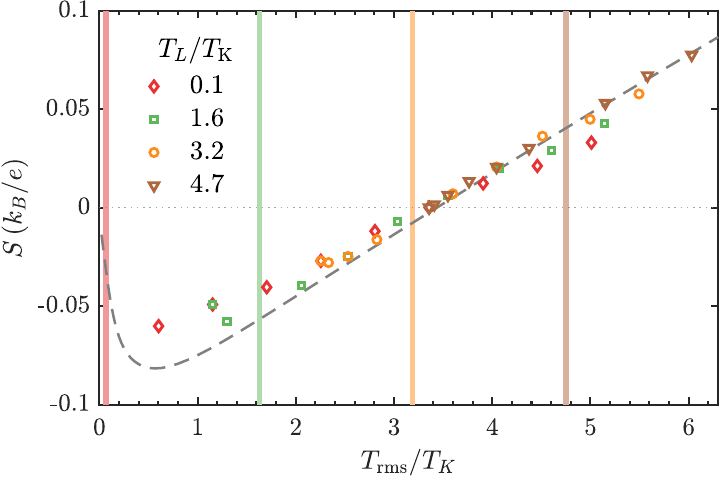}
 	\caption{The Seebeck coefficient $S\, (T_L,T_R)$ (colored symbols) with a
 	fixed $T_L$ (vertical colored lines) plotted against the effective temperature $T_{\rm rms}$.
 	The dashed line shows the equilibrium NRG data for $S_{0}\, (T_L=T_R=\Trms)$.
    \label{fig:SRMS}}
    \end{figure}

The comparison of $S(T_L,T_R)$ rescaled by the effective temperature $\Trms$ and the
linear response $S_0(T)$ from NRG is presented in \Fig{fig:SRMS}.
Unlike the differential conductance $G(\Trms)$, the rescaled $S(\Trms)$ data do not
fully resemble the linear response $S_0(T)$ behavior,
with increasing deviations for large temperature gradients.
We observe that the deviation of $S(\Trms)$ depends on the minima of the linear response $S_0$.
We define $T_p$ as the temperature, at which $S_0$ has a negative peak.
When the cold lead temperature is larger than $T_p$, $S(\Trms)$ lies
closer to the linear response $S_0$.
But for the case of a cold lead temperature below $T_p$,
left lead temperature
$T_L \approx 0.1\,\TK$ in our case
[cf. red diamonds in \Fig{fig:SRMS}], $S(\Trms)$ shows the largest deviations from the linear response $S_0$.

From the data in \Fig{fig:SRMS} we can conclude
that
the magnitude of the Seebeck coefficient is not enhanced
when compared to linear response $S_{\rm lin}$ under zero-bias conditions even with nonlinear temperature gradients.
Furthermore, the data in \Fig{fig:SRMS} shows
rather small values $|S|\lesssim 1$ for the Seebeck coefficient
in the Kondo regime.
This is in contrast, for example, to \Fig{fig:epsd}
where the Seebeck coefficient can reach values
an order of magnitude higher $|S|\lesssim 1$
just outside the local moment regime.
Based on these findings, let us briefly comment 
here on how to potentially enhance  the
thermoelectric response in the Kondo regime \cite{Sanchez2016}.
It was suggested that
an asymmetric coupling to the leads together with
a finite potential bias 
can improve the 
thermoelectric response in the Kondo regime,
as suggested in Ref.~\cite{PerezDaroca2018Apr}. While the NRG-tDMRG method is well-suited to handle such systems, 
a thorough investigation of this scenario
necessitates a detailed study
of its own and thus is
beyond the scope of the present work.

\begin{figure}[h]
 	\includegraphics[width=0.9\columnwidth]{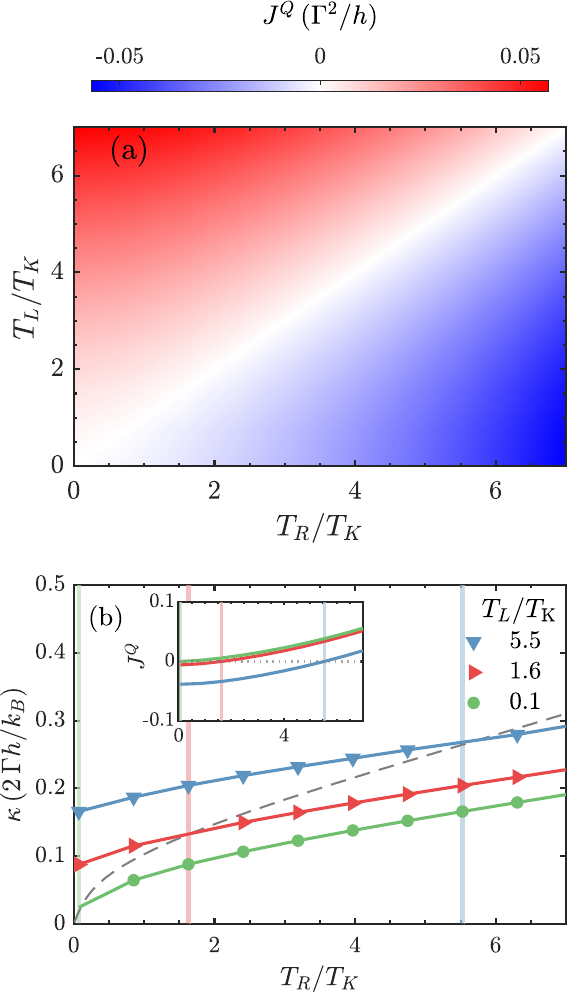}
 	\caption{
 	(a) The heat/energy current $J^Q$
	through a quantum dot [SIAM using \eqref{Eq:param}]
	as a function of the left lead temperature $T_L$ and the right lead temperature $T_R$.
	(b) shows the heat conductance $\kappa$ for different values of $T_L$, as indicated.
	The dashed line corresponds to the equilibrium NRG results for $\kappa_{0} \, (T=T_L=T_R)$.
	The inset presents the horizontal cross sections of panel (a) used for the
    estimation of $\kappa$ in panel (b).
\label{fig:JQ}}
\end{figure}

Lastly, we analyze the heat current and
heat conductance
in the presence of a finite temperature gradient. 
The heat current $J^Q$ across the quantum dot coupled to leads
with temperatures $T_L$ and $T_R$ is shown in \Fig{fig:JQ}(a).
Unlike the Seebeck coefficient, there exist no sign change in the heat conductance
characterizing the Kondo resonance. Thus, the heat current shows only
one sign change corresponding to the change in the sign of the temperature gradient $T_L-T_R$.
The electronic contribution to the heat conductance $\kappa$
calculated for the cross sections in panel (a) is presented in panel (b). 
It can be seen that for a constant $T_L$, 
$\kappa$ is enhanced with increase in $T_R$.
When reaching $T_R=T_L$, the heat conductance smoothly crosses the linear response $\kappa_{0}$.

\section{\label{Sec:Summary}Summary}

In this work we have provided accurate quantitative results for the thermoelectric 
transport properties of a Kondo-correlated quantum dot subject to nonlinear temperature and voltage gradients.
The calculations have been performed with the aid of numerical renormalization group--time-dependent
density matrix renormalization group method.
First of all, we have demonstrated that the thermoelectric behavior of the system,
involving charge and heat currents as well as the Seebeck coefficient,
can be qualitatively described by an effective global temperature $\Trms$.
Moreover, a detailed investigation of the zero-bias conductance with respect to the individual lead temperatures 
unveiled the Kondo regime as a circle in the plane of left-right lead temperatures,
further affirming the qualitative agreement with $\Trms$.  
The thermoelectric current also showed characteristic sign changes crossing over to the Kondo regime,
as a slightly distorted circle with the deviations occurring at large temperature gradients.
Moreover, we have provided a qualitative expression to fit the thermoelectric current
as a function of the left and right lead temperatures.
Finally, we have discussed the heat current and conductance near the Kondo regime,
which were mostly determined by the contribution from charge fluctuations,
hardly revealing characteristics of the Kondo resonance.

The thermoelectrics in the presence of finite
temperature gradients
at zero bias voltage
did not show any enhancement of
the thermoelectric properties originating from the nonlinear contributions in the Kondo regime. However, investigating the nonequilibrium
regime of asymmetrically coupled Kondo-correlated systems 
\cite{PerezDaroca2018Apr}
is a promising direction where NRG-tDMRG can yield reliable insights. This complex scenario warrants a dedicated study
of its own which goes beyond the scope of the
present
work and thus is left for the future.

\section{Acknowledgements}
This work was supported by the Polish National Science Centre from funds awarded through Decision No. NCN 2022/45/B/ST3/02826. 
A.M. acknowledges Brookhaven National Laboratory for hosting a research visit that contributed significantly to this work and is grateful to the NAWA-STER program for financial support provided for the visit through Decision no. PPI/STE2020/1/00007/U/00001.
A.W. was supported by the U.S. Department of Energy, Office of Science, Basic Energy Sciences, Materials Sciences and Engineering Division. This work was funded in part by the
Deutsche Forschungsgemeinschaft under Germany’s Excellence Strategy EXC-2111 (Project No. 390814868). We acknowledge helpful discussions with Kacper Wrze\`sniewski during the development of this work.

\section{Data availability Statement}

The datasets generated and analyzed for this work are publicly available on Zenodo at \href{https://doi.org/10.5281/zenodo.13773063}{https://doi.org/10.5281/zenodo.13773063}.

\appendix

\section{The Hybrid NRG-tDMRG method}
\label{App:NRG-tDMRG}

We use a hybrid NRG-tDMRG method to study the nonequilibrium
dynamics of the quantum dot coupled to leads with finite thermal and potential bias.
Below, we provide more details about this method and its extension
to finite thermal gradients.

\subsection{Hybrid discretization scheme}

Primarily, we separate the conduction band into modes that can be treated in equilibrium
and  out-of-equilibrium. i.e., the modes with $f_L(\omega) - f_R(\omega) = 0$
correspond to the modes that are at equilibrium
and $f_L(\omega) - f_R(\omega) \neq 0$ are the modes that are out of equilibrium,
where $f_\alpha(\omega)$ is the Fermi function for the lead $\alpha$.
For simplicity, we keep the largest $|\omega|$ that satisfies $f_L(\omega) - f_R(\omega) \neq 0$
as our effective bandwidth $D^*$ and define the transport window as $[-D^*,D^*]$
(essentially including more equilibrium modes into the tDMRG
part, which is easier to handle and provides a more accurate description
than moving more nonequilibrium modes into the NRG part).
The energies outside $|D^*|$ are discretized logarithmically
according to the discretization parameter $\Lambda$ and the energies inside $|D^*|$
are discretized linearly according to the discretization parameter $\delta$.
In this discretized setting, the coupling between the quantum dot energy level
$\e_d$ to a discretized mode in the lead $\alpha$ with momentum $k$
is given as $v_q = \sqrt{\Gamma_{\alpha} \delta_k/\pi}$,
where $\delta_k$ is the size of the corresponding interval in the discretized band. 

\subsection{Thermofield treatment}

We go on to describe the modes in this log-lin discretized band
under a thermofield description. This entails the introduction
of an auxiliary decoupled Hilbert space akin to the physical Hilbert space.
For a mode $c_{q1}$ in the physical Hilbert space,
where $q \equiv {\alpha,k,\sigma}$ is a composite index,
we introduce an auxiliary mode $c_{q2}$, where the index $2$ denotes
that the mode is in the auxiliary Hilbert space.
This enlarged Hilbert space is rotated by,
\be
\begin{pmatrix}
\tilde{c}_{q1}\\
\tilde{c}_{q2}
\end{pmatrix}
=
\begin{pmatrix}
\sqrt{1-f_q} & -\sqrt{f_q}\\
\sqrt{f_q} & \sqrt{1-f_q}
\end{pmatrix}
\,
\begin{pmatrix}
c_{q1}\\
c_{q2}
\end{pmatrix},
\ee
such that in the rotated tilde Hilbert space,
the modes $\tilde{c}_{q1}\, |\Omega\rangle = \tilde{c}_{q2}^{\dagger}\, |\Omega\rangle = 0$
can be interpreted as holes ($1$) and particles ($2$),
where $\Omega = \prod_q (\sqrt{1-f_q}\,|0,1\rangle_q + \sqrt{f_q}\,|1,0\rangle_q)$
is a pure state that can represent the thermal expectation
value of an operator $A$ on the physical lead as
$\langle A \rangle = \langle \Omega | A | \Omega \rangle$.

In the rotated Hilbert space, the lead Hamiltonian becomes,
\be
\mathcal{H}_{\rm lead} = H_{\rm lead} + H_{\rm aux} = \sum_{qj}\e_q c^{\dagger}_{qj} c_{qj} =  \sum_{qj} \e_q \tilde{c}^{\dagger}_{qj} \tilde{c}_{qj}.
\ee
We set $\e_{q2} = \e_{q1}$ to keep the total lead Hamiltonian $\mathcal{H}_{\rm lead}$ diagonal.
Similarly, the tunneling Hamiltonian in the rotated Hilbert space can be described as,
\be
H_{\rm tun} = \sum_{qj} (\tilde{v}_{qj}d^{\dagger}_{\alpha \s} \tilde{c}_{qj} + \rm H.c.),
\ee
where the couplings $\tilde{v}_{q1}=v_q \sqrt{1-f_q}$ and $\tilde{v}_{q2}=v_q \sqrt{f_q}$
become functions of the Fermi-Dirac distribution functions and, thus,
encompass the information about the nonequilibrium parameters,
such as the temperature and potential bias on the leads. 

\subsection{Recombination of the leads and tridiagonalization}

Outside the transport window $[-D^*,D^*]$,
the impurity is coupled to only half of the lead modes.
Since, $f_{\alpha}\to  1$ results in the hole coupling $\tilde{v}_{q1} \to 0$
and $f_{\alpha} \to 0$ results in the particle coupling $\tilde{v}_{q2} \to 0$.
This essentially means that both the high energy particle modes
and the low energy hole modes decouple from the impurity.
Whereas for the energies inside the transport window,
we use a different approach to simplify the structure.
Then, a single impurity coupled to two leads can be described using
an effective model with an impurity coupled to a single recombined lead
and such a recombination of the leads results in half of the modes
being decoupled from the system.
This results in the quantum impurity being coupled
to a set of hole lead modes and another set of particle lead modes.
In next step, we proceed to tridiagonalize these particle and lead modes separately,
resulting in two chains that are coupled to the impurity,
one from the hole modes and another from the particle modes.
In these chains, we can identify two sectors,
the sector from the high energy modes that lies closest to the impurity on the chain exhibiting properties of a Wilson chain,
i.e., energy scale separation and couplings that decay as $t_n \sim \Lambda^{-n}$.

\subsection{\label{App:Evolution}NRG treatment of high energy modes and time evolution}

Since the modes outside the transport window are essentially in equilibrium,
we recombine the holes and particles in the high energy sector for a more physically accurate description.
This results in our impurity being coupled to an effective Wilson chain
corresponding to the high energy sector, which is then further coupled to the separate hole and particle chains.
We treat the recombined high energy modes using the numerical renormalization group method
and extract the ground state of the high energy sector as $\ket{\phi_{\rm ini}}$.
$\ket{\phi_{\rm ini}}$ will act as the initial state for the high energy part of the chain,
where the low energy hole modes are kept empty and the particle modes filled.
Thus, our initial state for the time evolution $\ket{\psi_{\rm ini}}$ becomes
\be
\ket{\psi_{\rm ini}} = \ket{0} \otimes \ket{0} \otimes \dots \ket{0} \otimes \ket{\phi_{\rm ini}} \otimes \ket{1} \otimes \dots \otimes \ket{1}\otimes \ket{1}.
\ee
We time evolve $\ket{\psi_{\rm ini}}$ using
the second-order Trotter time evolution with a quench on the coupling
between the high energy and low energy sector over a finite time window.

The NRG-tDMRG calculations for the SIAM in this paper are performed with parameters 
$\Lambda=2.5$, 
$\delta/D^*=0.01$, $N_{\rm keep}=900$ kept states in the effective NRG basis of the renormalized impurity,
and a truncation tolerance of $\epsilon_{\rm SVD} = 10^{-5}$ for 
the tDMRG sweeps.
The observables are calculated for $100$ tDMRG sweeps with the first $20$ sweeps dedicated for the quench.
\subsection{Charge and heat current}

The particle current or the charge current $J_{\alpha}$
from the lead $\alpha$ to the quantum dot can be described as,
\bea
J_{\alpha \s} = e\, \langle \dot{N}_{\s}\rangle =-\frac{i}{\hbar} \langle [N_{\s},H]\rangle \nonumber \\
J_{\alpha \s}= \frac{e}{\hbar} \sum_{k} {\rm Im} \, (v_{\alpha k}\, \langle d^{\dagger} c_{\alpha k} \rangle )\\
\equiv \frac{e}{\hbar} \sum_{k}\sum_j {\rm Im}\, (\tilde{v}_{qj}\, \langle d^{\dagger} \tilde{c}_{qj} \rangle ).
\eea

Similarly, the energy current $J^{E}_{\alpha}$ from the lead $\alpha$
to the quantum dot can be described based on the lead Hamiltonian as,
\bea
J^E_{\alpha} = \langle \dot{H}_{\alpha} \rangle = -\frac{i}{\hbar} \langle[H_{\alpha},H]\rangle \nonumber \\
= \frac{2e}{\hbar} \sum_{k\s} \sum_j\e_q {\rm Im} (\tilde{v}_{qj} \langle d^{\dagger} \tilde{c}_{qj}\rangle).
\eea
The symmetrized current $J_{\s}(t)$ converges faster than the individual lead currents $J_{\alpha \s}$,
\be
J_{\s} (t) = \frac{1}{2} (J_{L\s} (t) - J_{R\s} (t))
\ee

\section{Extracting steady state observables via linear  prediction}
\label{App:prediction}

\begin{figure*}[t!]
\centering
\includegraphics[width=0.99\textwidth]{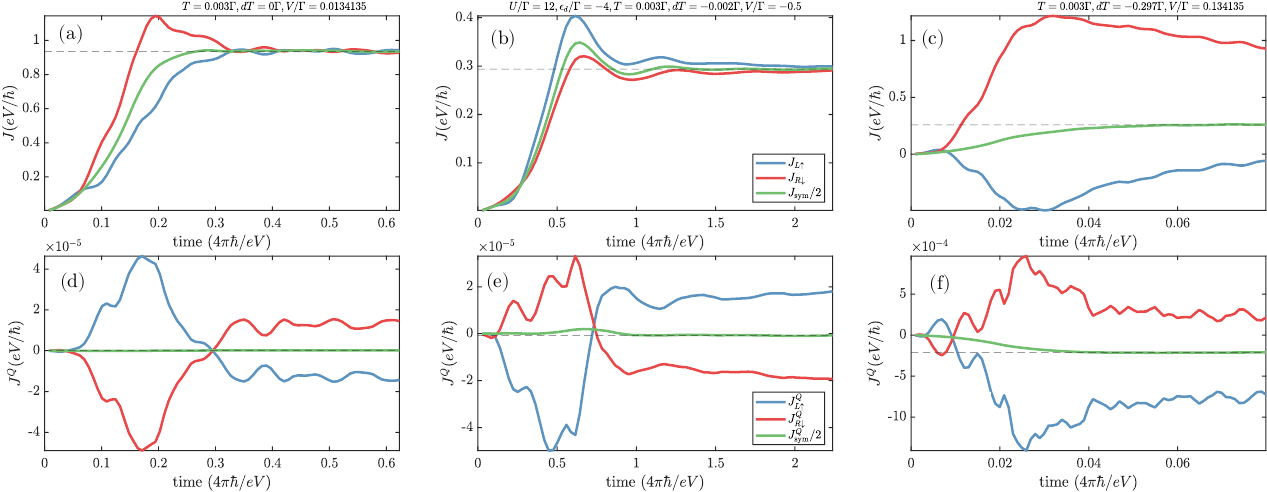}
\caption{
	The finite time dynamics of (a,b,c) charge current $J$ and (d,e,f) heat current $J^Q$ across a SIAM for
    representative values of the applied potential and temperature gradients.
    The horizontal dashed line shows the steady state value obtained from linear prediction.
\label{fig:pred}
}
\end{figure*}

The particle current shows a transient behavior during the quench window
and starts to oscillate around a steady-state value.
This steady state is extracted using linear regression.
We start by generating a kernel for the oscillating part based on the training window
\be
{y}_{n+1}
=
\underbrace{
\begin{bmatrix}
 a_{1} & a_{2} & \cdots & a_{n}\\
\end{bmatrix}
}_{K}
\,
\begin{bmatrix}
x_1  \\
x_2 \\
\vdots \\
x_n
\end{bmatrix},
\ee
where the kernel $K$ estimates the next data point $y_{n+1}$
based on the previous $n$ data points $\{x_1,x_2 \cdots x_n\}$.
We estimate $K$ as the least squared approximation of the data points in the training window.
The spectral decomposition of the kernel has the information
about the oscillating behavior of the data.
In particular, we isolate the eigenvector with the real eigenvalue
(corresponding to the non-oscillating part) to estimate the steady state current at $t \to \infty$
\be
J(t \to \infty) = \frac{\lVert \vec{e}_0 \rVert}{\sqrt{e_0}},
\ee
where $\vec{e}_0$ is the eigenvector corresponding to the real eigenvalue $e_0$.
Figure~\ref{fig:pred} shows the charge current (a,b,c) and the heat current (d,e,f)
dynamics of a SIAM using \eqref{Eq:param} obtained from NRG-tDMRG. 
The steady state value estimated from linear prediction is shown as the 
horizontal dashed line.

\section{\label{App:TK} Effective Kondo energy scale}

\begin{figure}[t]
	\centering
 	\includegraphics[width=0.99\columnwidth]{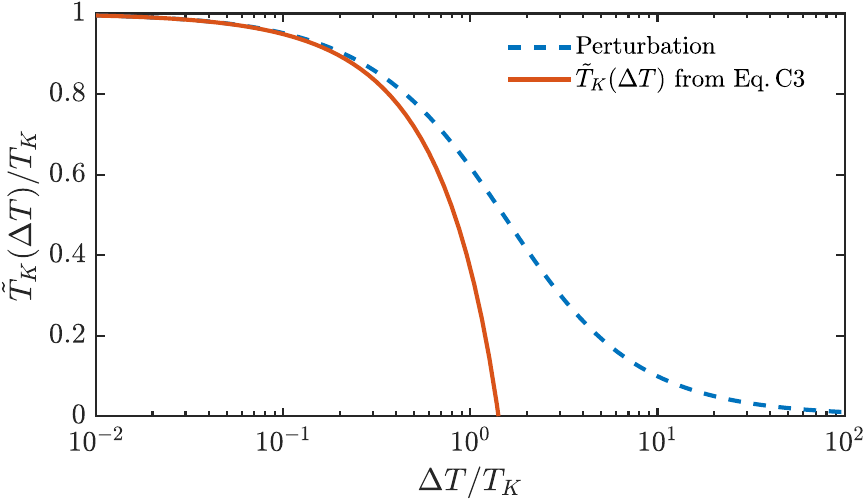}
 	\caption{
	The Kondo energy scale in \Eq{TK_deltaT}
	as a function of the temperature gradient $\Delta T$
	compared to the perturbation theory (\Eq{PT}) results from Ref.~\cite{Sierra2017Aug}.
    \label{fig:TK}}
    \end{figure}   

Analytical dependence of the Kondo energy scale
on the temperature gradient has been discussed in Ref. ~\cite{Sierra2017Aug}
by using the perturbation theory and slave-boson mean-field theory.
According to the perturbation theory, the Kondo energy scale depends on the temperature gradient as,
\be
\tilde{T}^{\rm PT}_{\rm {K}} (\Delta T) = \sqrt{ T^2_{\rm K} + \left(\tfrac{\Delta T}{2}\right)^2} -\tfrac{\Delta T}{2},
\label{Eq:PT}
\ee
where $\tilde{T}_{\rm K}$ is defined as the energy scale at which 
the second-order term dominates in the perturbation expansion of 
the conductance in the Kondo model [Eq. (11) and Eq. (13) from the Ref.\cite{Sierra2017Aug}].
Nevertheless, throughout this paper, $\TK$ denotes the intrinsic Kondo temperature
of the system as defined in \Eq{Haldane}.

From the NRG-tDMRG calculations, an effective temperature of
$\Trms=T_{\rm K}$ in the $T_L-T_R$ plane represents a circle of the form $T_L^2 + T_R^2 = 2\, T_{\rm K}^2$.
To compare with the results from perturbation theory,
we consider $T_L=T$ and $T_R = T+\Delta T$.
Thus, we can define the energy scale
$\tilde{T}_{\rm K} (\Delta T)$ for a fixed $\Delta T$,
\be
  \bigl. G(T, T + \Delta T)\bigr|_{T = \tilde{T}_{\rm K}}
  = G_0/2,
\ ,\label{Eq:TKtilde}
\ee
i.e., as the temperature $T$ at which $G(T,T+\Delta T)$
reaches the half maximum of the conductance peak $G_0$
at $T=\Delta T=0$. By definition then,
$\tilde{T}_{\rm K}$ reduces with increasing $\Delta T$
towards zero, and becomes undefined for sufficiently large $\Delta T > \TK$
once $G(T,T+\Delta T) < G_0/2$ for all $T$.
In this sense,  $\tilde{T}_{\rm K} \to 0$ does not indicate a small physical Kondo scale, per se,
but rather the disappearance of the Kondo physics.
Based on the Kondo circle $T_L^2 + T_R^2 = 2\, T_{\rm K}^2$
[cf. \Sec{G}], \Eq{TKtilde}
provides an analytical expression for $\tilde{T}_{\rm {K}}$
\be
\tilde{T}_{\rm {K}} (\Delta T) = \sqrt{ T^2_{\rm K} - \left(\tfrac{\Delta T}{2}\right)^2} -\tfrac{\Delta T}{2}.
\label{Eq:TK_deltaT}
\ee
This expression for $\tilde{T}_{\rm {K}} (\Delta T)$ is very similar to the perturbation theory result,
except for the difference in sign of the $\Delta T^2$ term under the square root. 

The temperature $\tilde{T}_{\rm K}$ defined on the Kondo circle and
the Kondo temperature $\tilde{T}_{\rm K}^{\rm PT}$ from the perturbation theory
show good agreement for small $\Delta T$ [cf.~\Fig{fig:TK}].
With increasing $\Delta T$, $\tilde{T}_{\rm K}$ decays faster than
$\tilde{T}_{\rm K}^{\rm PT}$ and proceed to become undefined beyond $\Delta T = \sqrt{2} \,T_K$.

%


\end{document}